\documentclass[12pt]{iopart}

\usepackage{iopams}
\usepackage{graphicx}
\begin{document}

\title[]{High-field magnetization of a two dimensional spin frustration system
  Ni$_{5}$(TeO$_{3}$)$_{4}$X$_{2}$ (X = Br and Cl)}

\author{J. L. Her$^{1}$, Y. H. Matsuda$^{1}$, K. Suga$^{1}$, K. Kindo$^{1}$, S. Takeyama$^{1}$, H. Berger$^{2}$, and H. D. Yang$^{3}$}
\address{$^{1}$ Institute for Solid State Physics, University of Tokyo, Japan}
\address{$^{2}$ Institutes of Physics of Complex Matter, EPFL, Lausanne, Switzerland}
\address{$^{3}$ Department of Physics, Center for Nano Science and Nano Technology, National Sun
Yat-sen University, Taiwan}
\begin{abstract}
High-field magnetization, M($H$), on Ni$_{5}$(TeO$_{3}$)$_{4}$X$_{2}$ (X = Br and Cl) were measured by using a pulse magnet. These compounds have a two dimensional crystal structure and a distorted kagome spin frustrated system which is builded by the Ni$^{2+}$ ions ($\textbf{S}$ = 1). The N\'{e}el transition temperatures are $T_{N}~\sim$ 28 and 23 K for X = Br and Cl, respectively. When $T~<~T_{N}$, we observed a step-like transition at $H_{c}~\sim$ 11 and 10 T for X = Br and Cl, respectively. On the other hand, at $T~>~T_{N}$, the field-dependent magnetization curves behaved like a monotonically increasing straight line up to 55 T. The $H_{c}$ value is close to those obtained by previous spin resonance studies in which a model of spin-flop scenario was proposed to explain the field-dependent resonance spectra. Their model predicts a further transition at around 23 T, however, our observations did not show any plateau behaviors, saturation or other anomalies up to 55 T, suggesting that the further transition possibly exists at a much higher field region.

\end{abstract}

\maketitle

\section{\label{sec:level1}Introduction}

The frustration of spins coupled antiferromagnetically is an ongoing and interesting subject in condensed matter physics. The quantum spin fluctuation in such a frustrated system, which causes large amount of degenerating ground states, interferes with formation of a long-range N\'{e}el order. On the other hand, the magnetic anisotropy of spins can open a gap in low lying exciting spectrum, leading to the occurrence of long-range ordering states. The competition between magnetic anisotropy and spin frustration results in various kinds of magnetic ordering phases. Two-dimensional (2-D) kagome spin frustrated system is a remarkable target to study these interesting magnetic phases, due to the anisotropic nature and unique spin arrangement. Extensive studies were performed on the kagome spin system and discovered diverse magnetic ground states, such as quantum liquid\cite{JACS.127.13462}, spin gap\cite{JPSJ.77.043707}, antiferromagnetic\cite{CM.15.68}, ferromagnetic states\cite{JMC.11.1152} and so on.

Ni$_{5}$(TeO$_{3}$)$_{4}$X$_{2}$ (X = Br and Cl) is a new series compound of the kagome spin system, having a well separated 2-D layer structure in which the [Ni$_{5}$O$_{17}$X$_{2}$] units constructed a 2-D layer and the layers separated by the Coulomb repulsion of the lone pairs of Te$^{4+}$ ions\cite{CM.15.68}. The Ni$^{2+}$ ions serve as magnetic centers with $S=1$, and couple to each other by an antiferromagnetic superexchange interaction\cite{CM.15.68}. The long-range N\'{e}el ordering temperatures are 28 and 23 K for X = Br and Cl, respectively\cite{CM.15.68}.  The anisotropic properties were investigated on single crystal Ni$_{5}$(TeO$_{3}$)$_{4}$Br$_{2}$, showing the g$_{\parallel}$=2.45 and g$_{\perp}$=2.53\cite{JPCM:145278}. Recently, the noncollinear arrangements of the Ni sublattices of Ni$_{5}$(TeO$_{3}$)$_{4}$Br$_{2}$ were observed by neutron diffraction and magnetization measurements\cite{pregelj:144408}. These studies reveal very complicated spin interactions and a unique ground state. The magnetic field effects of these compounds have been studied by high-field electron spin resonance (ESR) experiments, in which the antiferromagnetic resonance modes were observed on both X = Br and Cl compounds\cite{JPCM:145278,pregelj:144408,Arcon2007e349,Pregelj2008950}. The observed lowest resonance mode is first softened and then hardened by a magnetic field, having a critical field $\sim$10.7 and 10 T for X = Br and Cl, respectively\cite{pregelj:144408,mihaly:174403}, suggesting that a spin-flop-like transition exists. However, the models proposed by these two reports have huge difference on handling the spin isotropic effect, in which the spin anisotropy was regarded as important in one report\cite{pregelj:144408} whilst neglected in the other\cite{mihaly:174403}. The high-field magnetization measurements can provide more information of high-field state of these samples. Very recently, Pregelj $et~al$. reported a magnetization study of Ni$_{5}$(TeO$_{3}$)$_{4}$Br$_{2}$ in magnetic fields up to 12 T. A transition peak was observed in dM/d$H$ curve at $\sim$ 11 T, which is suggested to be related to the spin-flop-like transition in ESR experiments\cite{Pregelj064407}. A model, including spin-ion anisotropy, was proposed to explain this observation. In addition, this model predicts that a second transition occurs at 23 T. It is interesting to measure the magnetization in higher fields, to confirm if there is another transition.

In the present study, we focus on the field-dependent magnetization of these compounds by using pulse magnet which generates magnetic fields up to 55 T. A step-like transition, at around 11 T(10 T) for X = Br(Cl), was observed in M($H$) curves which is consistent with Pregelj's results and also with ESR experiments\cite{pregelj:144408,mihaly:174403,Pregelj064407}. Interestingly, there is no signature of other field-induced transition up to 55 T. However, as far as the M($H$) curves showed unsaturated behavior up to the highest field in the current measurements, there exists a possibility of the second transition occurring at fields higher than 55 T.

\section{Experiment}
\label{Experi}
A plate-like single crystal was used in the measurement. The details of preparation were described in a previous report\cite{pregelj:144408}. High-field magnetization measurement, M($H$), was performed by an induction method using a pulse magnet. This system can generate pulse-fields up to 55 T and the duration time is 40 ms. To obtain stronger signal, several pieces of single crystal were used, which have the total mass $\sim$ 38 and 17 mg for X = Br and Cl, respectively. Magnetic field is applied perpendicular to the crystalline surface, i.e. the magnetic field is oriented along the a*-axis, where a* denotes the normal direction of 2-D layers. The absolute values of magnetization curves are carefully calibrated by low-field magnetization measurements which were performed by SQUID magtometer(Quantum Design MPMS).
\section{Results and Discussion}
\label{Res}
\begin{figure}
  \begin{center}
  \includegraphics[width=3in]{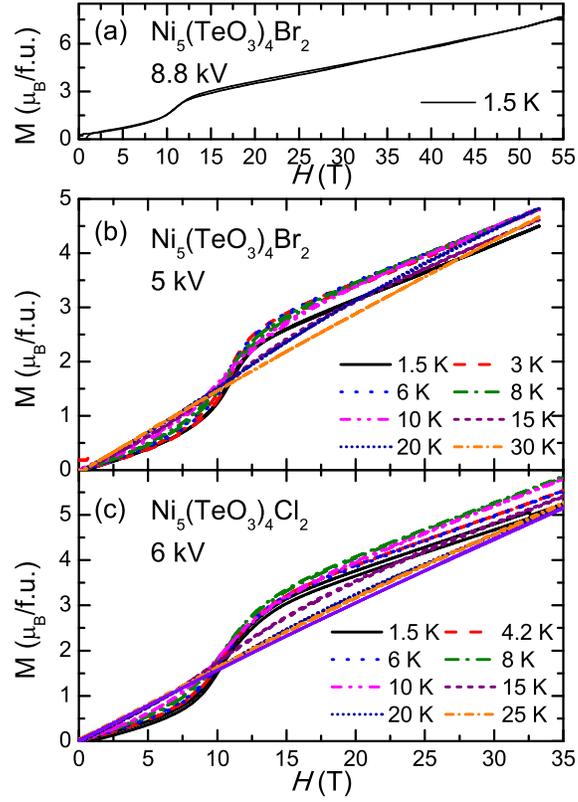}
  \end{center}
  \caption{(Color online)(a) M(\textit{H}) of Ni$_{5}$(TeO$_{3}$)$_{4}$Br$_{2}$ at 1.5 K. (b)(c) M(\textit{H}) curves of Ni$_{5}$(TeO$_{3}$)$_{4}$Br$_{2}$ and Ni$_{5}$(TeO$_{3}$)$_{4}$Cl$_{2}$ at some selected temperatures.}\label{mh}
\end{figure}
\begin{figure}
  %Requires \usepackage{graphicx}
  \begin{center}
  \includegraphics[width=5in]{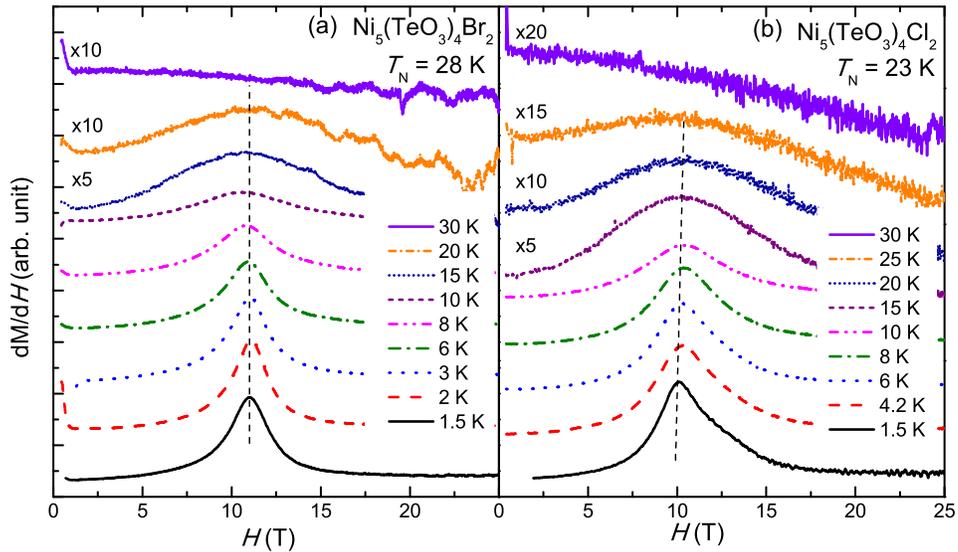}
  \end{center}
  \caption{(Color online)dM/d\textit{H} curves of Ni$_{5}$(TeO$_{3}$)$_{4}$Br$_{2}$ and Ni$_{5}$(TeO$_{3}$)$_{4}$Cl$_{2}$ at some selected temperatures.}\label{dmdh}
\end{figure}
Figure \ref{mh}(a) shows the M($H$) curve of Ni$_{5}$(TeO$_{3}$)$_{4}$Br$_{2}$ at 1.5 K. A step-like transition can be clearly seen at $H_{c}\sim$ 11 T. When $H~>~H_{c}$, the M($H$) curve shows a monotonic linear increase, which shows neither saturation nor a plateau behavior up to $H~\sim$ 55 T. In Fig. \ref{mh}(b) and (c), we show the M($H$) curves at different temperatures on X = Br and Cl compounds. It is found that two compounds show a very similar behavior. The M($H$) curve at 30 K(25 K) of a Br(Cl) sample is almost a straight line which represents the paramagnetic property at $T > T_{N}$. When temperatures are lower than $T_{N}$, the step-like transitions start to appear at around 10 T and become more and more prominent at lower temperatures in both samples. In addition, M($H$) curves of increasing and decreasing field coincide each other without showing any hysteresis in all the case. The $H_{c}$ values are consistent with ESR results\cite{pregelj:144408,mihaly:174403}, and also the previous field-dependent magnetization data\cite{Pregelj064407}.

In order to assist clearance of the transition point, we show the dM/d$H$ curves in Fig. \ref{dmdh}. At the lowest temperature, there is a sharp peak which is related to a step-like transition. This peak becomes broader with increasing temperature, and disappears at high temperatures. Although the behaviors of two samples are quite similar, there are some differences between these two samples. First, the peak disappears just above N\'{e}el Temperature for X = Br sample. However, there  remains a very weak and broad peak at $T > T_{N}$ (i.e. 25 K) for X = Cl sample. It is possibly related to some short-range ordering component which presents at $T > T_{N}$. Second, in the case of X = Cl, the peak positions, $H_{c}$, are slightly shifted toward higher fields with increasing temperature. However, the peak positions are almost the same in the case of X = Br, indicating that the temperature dependence of this transition is very weak.

\begin{figure}
  \begin{center}
  \includegraphics[width=3in]{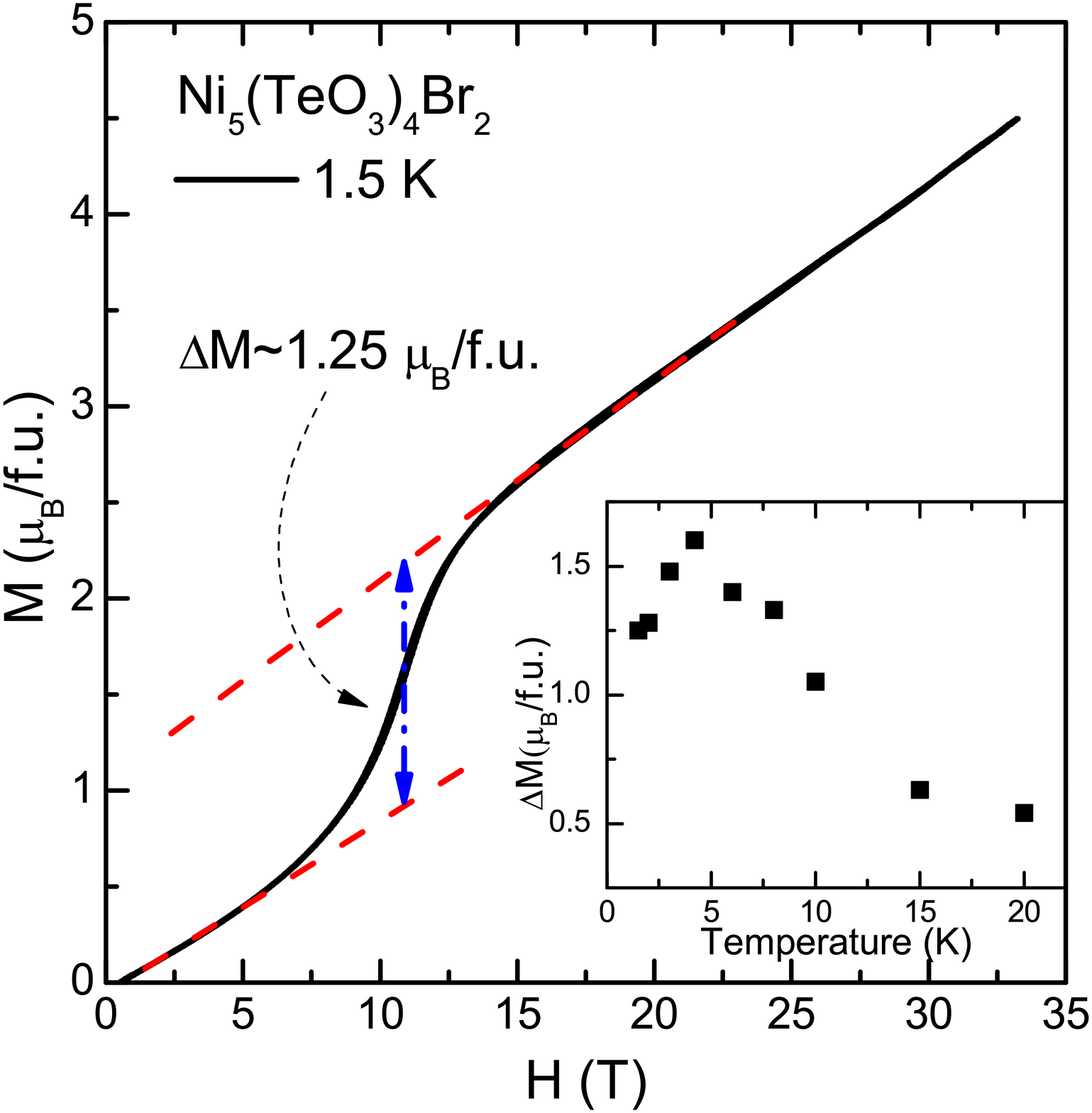}
  \end{center}
  \caption{(Color online)Typical plot for determining $\Delta$M. The red dash lines is a guiding by eye. Inset shows the temperature-dependent $\Delta$M.}\label{step}
\end{figure}

The difference of magnetization between low-field and high-field states at the transition ($\Delta$M) can be roughly determined by the M($H$) curves. In figure \ref{step}, we show a typical example of X = Br sample at 1.5 K, in which the $\Delta$M is $\sim$ 1.25 $\mu_{B}$ per formula. The $\Delta$M value increases with decreasing temperature and slightly decreases below 5 K (the inset of figure \ref{step}). Interestingly, the $\Delta$M value at lowest temperature is nearly one half of the magnetic moment of Ni$^{2+}$ ions , $m_{Ni}~\sim~2.53~\mu_{B}$\cite{JPCM:145278}, indicating that only a part of Ni$^{2+}$ ion spins participated in the transition.

\begin{figure}
  \begin{center}
  \includegraphics[width=3in]{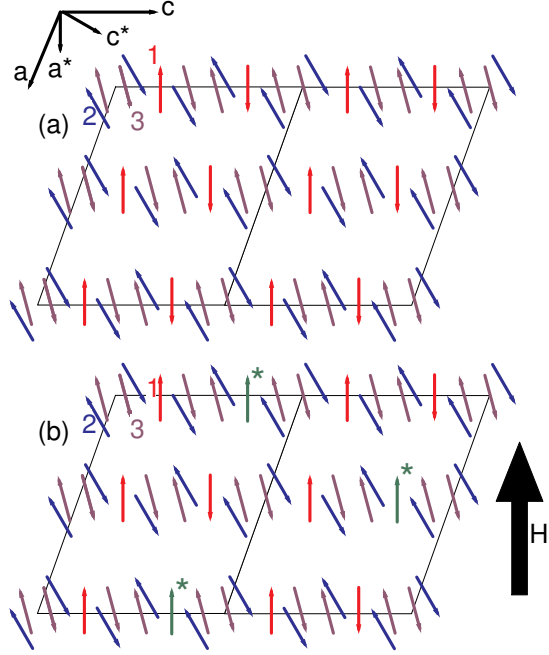}
  \end{center}
  \caption{(Color online)(a) Sketch of spin configuration at low-field of studied samples ($ac$ projection)\cite{pregelj:144408}. Red, purple, and royal arrows represent the spin orientation of Ni1, Ni2, and Ni3 ions. (b) Sketch of high-field spin configuration. The large arrow beside denotes the direction of applied magnetic field.}\label{model}
\end{figure}

According to the results of neutron and X-ray diffraction measurements, there are 20 Ni$^{2+}$ ions in an unit cell (Z-factor = 4). The Ni ions locate in three different crystallographic sites, in which Ni1 is in Wyckoff site 4e; Ni2 and Ni3 are in Wyckoff site 8f\cite{CM.15.68,pregelj:144408}. The spin orientations depend on the Ni sites. Figure\ref{model}(a) shows the sketch of the ten sub-lattice spin configuration of these compounds, in which 1, 2, and 3 denote Ni1, Ni2 and Ni3 site. We use the same notation as in the neutron scattering report\cite{pregelj:144408}. It should be noted that the spins of Ni1 ions are nearly parallel to a*-axis ($\theta$ $<$ 5$^{o}$) at $T < T_{N}$ which is also the direction of applying fields in our experiments. Due to this complicated spin configuration, it is possible that different spin sub-lattices respond differently to the applied magnetic field.

Based on this concept, we propose a simplified spin-flip model to explain our observed step-like transition. At low fields, there are four Ni1 spins in an unit cell, which are two upward and two downward spins. When $H~>~H_{c}$, one downward spin flips to upward and causes a moment change, $\Delta$M, by 2$m_{Ni}$ per unit cell. Since Z-factor is 4, the $\Delta$M per formula is one half of $m_{Ni}$, which is consistent with our results. A possible spin configuration of high-field state is showing in Fig. \ref{model}(b), where the flipped spin is denoted by a green arrow with a star (color online). In addition, the spin-flip transition usually shows a plateau at a high-field state, however our data shows a monotonically increasing feature. It is possibly caused by the field dependence of the spins of Ni2 and Ni3 ions, which are tilted toward the field-direction and produces the non-plateau high-field state. Furthermore, in the Fig. \ref{model}(b), there is still a downward Ni1 ion-spin in an unit cell, suggesting a second transition which correlated with another spin-flip behavior possibly exists at higher field over 55 T. This simplified spin-flip model can only explain the $\Delta$M of the low-temperature M($H$) curves. At higher temperatures, the thermal fluctuation or the complex out-of-plane spin arrangement might occur and reduce the $\Delta$M values.

In the recently report of Pregelj $et~al.$, the 11 T transition was assigned as a spin-flop-like transition from an in-plane antiferromagnetic phase to a complex out-of-plane spin arrangement. Our measurements were carried out only in a*-direction, therefore, we cannot observe the effect of the out-of-plane spin arrangement directly. They also predicted that the transition field will increase with increasing temperature. However, our results show that the temperature dependence of transition field is negligibly weak, but the $\Delta$M is strongly affected by a thermal fluctuation. Furthermore, they predict that a ferromagnetic ordering phase of [Ni$_{5}$O$_{17}$X$_{2}$] units will exist above 24 T. However, we do not observe the transition in the region $H_{c}<~H~<$ 55 T, indicating that the model proposed by Pregelj $et~al.$ requires further improvements.

\section{Summary}
\label{Sum}
High-field magnetization curves of Ni$_{5}$(TeO$_{3}$)$_{4}$X$_{2}$ (X = Br and Cl) were measured at different temperatures. Both samples showed similar field-dependent behavior in various temperatures. A step-like transition was observed at 11 T(10 T) for X = Br(Cl), which was consistent with other study by ESR spectroscopy or magnetization measurements. At around 24 T, we did not observe any sign of another transition, which was inconsistent with the prediction by Pregelj's report\cite{Pregelj064407}. From M($H$) curves, we found that $\Delta$M per formula is close to one half of $m_{Ni}$. We proposed a simplified spin-flip model to explain the step-like transition, which also suggested that a further transition could exist in the range higher than 55 T. Further magnetization measurements at higher fields and a more general spin-flop model are necessary.

\ack
This work was supported by Grant-in-Aid for Scientific Research on
priority Areas "High Field Spin Science in 100 T" (No. 451) from
the Ministry of Education, Culture, Sports, Science and Technology
(MEXT) of Japan.

\section*{References}
\bibliographystyle{iopart-num}
\bibliography{Ni5TeO34Br2}

\end{document}